\documentclass[reprint, amsmath,amssymb, aps, showpacs, superscriptaddress, pra]{revtex4-1}

\usepackage{graphicx}
\usepackage[breaklinks=true,colorlinks=true,linkcolor=blue,urlcolor=blue,citecolor=blue]{hyperref}

\def\bra#1{{\langle #1 |}}
\def\ket#1{{| #1 \rangle}}

\begin{document}
\preprint{}

\title{Quantum computation mediated by ancillary qudits and spin coherent states}
\author{Timothy J. Proctor}
 \email{py08tjp@leeds.ac.uk}
\affiliation{School of Physics and Astronomy, E C Stoner Building, University of Leeds, Leeds, LS2 9JT, UK}
\author{Shane Dooley}
\affiliation{School of Physics and Astronomy, E C Stoner Building, University of Leeds, Leeds, LS2 9JT, UK}
\date{\today}
\author{Viv Kendon}
 \thanks{current address: Department of Physics, Durham University, Durham, DH1 3LE, UK}
\affiliation{School of Physics and Astronomy, E C Stoner Building, University of Leeds, Leeds, LS2 9JT, UK}
\date{\today}

\begin{abstract}
Models of universal quantum computation in which the required interactions between register (computational) qubits are mediated by some ancillary system are highly relevant to experimental realisations of a quantum computer. We introduce such a universal model that employs a $d$-dimensional ancillary qudit. The ancilla-register interactions take the form of controlled displacements operators, with a displacement operator defined on the periodic and discrete lattice phase space of a qudit. We show that these interactions can implement controlled phase gates on the register by utilising geometric phases that are created when closed loops are traversed in this phase space. The extra degrees of freedom of the ancilla can be harnessed to reduce the number of operations required for certain gate sequences. In particular, we see that the computational advantages of the quantum bus (qubus) architecture, which employs a field-mode ancilla, are also applicable to this model. We then explore an alternative ancilla-mediated model which employs a spin-ensemble as the ancillary system and again the interactions with the register qubits are via controlled displacement operators, with a displacement operator defined on the Bloch sphere phase space of the spin coherent states of the ensemble. We discuss the computational advantages of this model and its relationship with the qubus architecture.
\end{abstract}


\maketitle

\section{Introduction}
A quantum computer has the potential to solve certain problems and implement simulations faster than any classical computer \cite{feynman1985quantum,shor1997polynomial}. Although many steps have been made towards a physical realisation of a quantum computer, a device that can outperform a classical computer remains a huge challenge. The original theoretical setting for quantum computation is the gate model \cite{feynman1985quantum, barenco1995elementary}, where a global unitary on a register of computational qubits is decomposed into some universal finite gate set, often composed of a single entangling two-qubit gate and a universal set of single-qubit unitaries \cite{bremner2002practical, brylinski2002universal}. However, this model requires both individual qubit addressability, to implement single-qubit unitaries on each register qubit, and controllable coherent two-qubit interactions between arbitrary pairs of register qubits. This can be very experimentally challenging and so, motivated by this, alternative models of quantum computation have been developed.
\newline
\indent
One possible route to improving the physical viability of a model is to mediate the multi-qubit gates between computational qubits using some ancillary system. The original ion trap gate of Cirac and Zoller is such a scheme, where the ancillary system in this case is the collective quantised motion of the ions \cite{cirac1995quantum}. We shall refer to computational architectures of this type as \emph{ancilla-mediated quantum computation} (AMQC). This encompasses many of the experimental demonstrations of quantum computation and AMQC has many advantages over a direct implementation of the gate model. Firstly, the ancillary system may be of a different physical type that is optimised for communication between isolated low decoherence qubits in a computational register. Such hybrid systems have been proposed or physically realised in a variety of physical setups, an example being the coupling of spin or atomic qubits via ancillary photonic qubits \cite{carter2013quantum,tiecke2014nanophotonic}. Indeed, models of universal quantum computation in which an ancillary qubit mediates \emph{all} the required operations on the register qubits via a single fixed-time interaction between the ancilla and a single register qubit at a time have been developed \cite{anders2010ancilla, kashefi2009twisted,shah2013ancilla,halil2014minimum,proctor2013universal,proctor2014minimal}. This is known as \emph{ancilla-driven} or  \emph{ancilla-controlled} quantum computation when measurements of the ancilla drive the evolution \cite{anders2010ancilla, kashefi2009twisted,shah2013ancilla,halil2014minimum} or when all of the dynamics are unitary \cite{proctor2013universal,proctor2014minimal} respectively.
\newline
\indent
However, in general, the mediating ancillary system need not be a qubit but may be of any dimension. This is the case in a variety of experimental settings such as, superconducting qubits coupled via a transmission line resonator \cite{,PhysRevLett.95.060501, majer2007coupling}, semiconductor spin qubits coupled optically \cite{yamamoto2009optically}, or the coupling of a Cooper-pair box with a micro-mechanical resonator \cite{armour2002entanglement}. A well studied computational model which harnesses a higher dimensional ancilla is \emph{quantum bus} (\emph{qubus}) computation \cite{milburn1999simulating,wang2002simulation,spiller2006quantum, brown2011ancilla, louis2007efficiencies,munro2005efficient,proctor2012low} which employs a field-mode `bus' and the interactions with the register qubits are via controlled displacements \cite{milburn1999simulating,wang2002simulation,spiller2006quantum,brown2011ancilla, louis2007efficiencies} or controlled rotations of the field-mode \cite{spiller2006quantum, munro2005efficient, proctor2012low}. The continuous-variable nature of the displacement operator for a field-mode can have additional advantages in terms of the computational power of the model. In particular, certain gate sequences can be implemented using fewer bus-qubit interactions than if each gate was implemented individually \cite{louis2007efficiencies,brown2011ancilla} and these techniques can be used to implement certain quantum circuits with a lower scaling in the total number of interactions required in comparison to the standard quantum circuit model \cite{noteE}.
\newline
\indent
A possible alternative ancillary system to a field-mode is a $d$-dimensional system, a \emph{qudit}. Models that utilise qudits have been shown to exhibit a reduction in the number of operations required to implement a Toffoli gate \cite{ralph2007efficient,borrelli2011simple,lanyon2009simplifying}. In particular, it has been shown that using a qudit ancilla can aid a computational model, with advantages including large savings in the number of operations required to implement a generalised Toffoli gate (a unitary controlled on multiple qubits) \cite{ionicioiu2009generalized} and simple methods for realising generalised parity measurements on a register of qubits \cite{ionicioiu2008generalized}. These results are not directly applicable to the qubus model, however we show, using the formalism for the finite lattice phase space of a qudit \cite{wootters1987wigner,vourdas2004quantum}, that the computational advantages of a field-mode bus also apply in the case of a qudit ancilla. We develop a full ancilla-mediated model of quantum computation based only on controlled displacement operators acting on an ancilla qudit. The previous work \cite{ionicioiu2009generalized,ionicioiu2008generalized} on ancillary qudits can also be understood within this framework and we show that the computational advantages demonstrated in the qubus model can be transferred into this finite dimensional context.
\newline
\indent
One possible physical realisation of a qudit is in the shared excitations of an ensemble of qubits (the Dicke states), with such ensembles realised and coherently manipulated using nitrogen-vacancy (NV) centers in diamond \cite{zhu2011coherent} and ensembles of caesium atoms \cite{christensen2013quantum}. However such systems are also naturally described using the language of the continuously parameterised spin-coherent states \cite{radcliffe1971some,arecchi1972atomic}. We further show that with an appropriately defined controlled displacement operator, based on rotations on a Bloch sphere, we can introduce an alternative ancilla-mediated model. Although individual two-qubit gates can be implemented in a simple manner, due to the spherical nature of the phase space the equivalent displacement sequences to those used in qubus computation to reduce the total number of interactions required do not implement the desired gates with perfect fidelity in this case. However, we show that these sequences exhibit negligible intrinsic error for spin-coherent states consisting of realistic numbers of spins. We begin with some essential definitions and a review of the field-mode qubus model.


\section{Background \label{fmqubus}}
\subsection{Definitions and phase space formalism}
We denote the Pauli operators for the $j^{th}$ qubit by $X_j$, $Y_j$ and $Z_j$  and the $+1$ and $-1$ eigenstates of $Z$ 
by $\ket{0}$ and $\ket{1}$ respectively (the computational basis). We define a general controlled gate, controlled by the $j^{th}$ qubit, by
\begin{equation} C^j_{k}(U,V):= \ket{0}\bra{0}_j \otimes U_k +  \ket{1}\bra{1}_j \otimes V_k ,\end{equation}
where $U$ and $V$ are unitary operators acting on the target system $k$ and $CU:=C(\mathbb{I},U)$. Furthermore, we take the standard definition for
the single-qubit phase gate
\begin{equation} R(\theta) = \ket{0}\bra{0} + e^{i \theta} \ket{1}\bra{1}. \end{equation}
Finally, we denote the set of integers modulo $d$ by $\mathbb{Z}(d)=\{0,1,...,d-1\}$ and the $d^{th}$ root of unity by $\omega_d$, using the notation
\begin{equation} \omega_d(a) := \omega_d^a = e^{i \frac{2 \pi a}{d}}. \end{equation}
For a field mode, translations in position and momentum are given by
\begin{equation} X(x):=\exp(-ix\hat{p}), \hspace{1cm} P(p):=\exp(ip\hat{x}), \end{equation}
respectively, where the position and momentum operators, $\hat{x}$ and $\hat{p}$, obey $[\hat{x},\hat{p}]=i$ ($\hbar=1$). Their commutation relation can be expressed in Weyl form as
\begin{equation} P(p)X(x)=e^{ixp}X(x)P(p). \label{Weyl} \end{equation} 
We then define the displacement operator by
\begin{equation}\mathcal{D}(x,p):=e^{-\frac{i}{2} xp} P(p)X(x), \label{displace} \end{equation}
which can be also be expressed as
\begin{equation} \mathcal{D}(x,p)= \exp (i(p \hat{x} - x \hat{p})),\label{entangledisp}\end{equation} 
using the Baker-Campbell-Hausdorff formula  \cite{gazeau2009coherent}. We then define the coherent states by
\begin{equation} \ket{x,p} := \mathcal{D}(x,p) \ket{\psi_0},\label{cohst} \end{equation}
where $\ket{\psi_0}$ is normally taken to be the vacuum. We have the identity
\begin{equation}\mathcal{D}(x_2,p_2)\mathcal{D}(x_1,p_1)= \exp(i \phi)\mathcal{D}(x_1+x_2,p_1+p_2), \label{comd} \end{equation}
where $\phi=(x_1p_2-p_1x_2)/2$ and hence a displacement operator $\mathcal{D}(x_1,p_1)$ translates the phase space point $(x_0,p_0)$ to the point $(x_0+x_1,p_0+p_1)$. A set of displacements that form a closed loop in phase space will create a geometric phase, given by $\exp(\pm i \mathcal{A})$ where $\mathcal{A}$ is the area enclosed \cite{wang2002simulation, spiller2006quantum} and with the sign dependent on the direction that the path is traversed. A simple case involves translations around a rectangle, given by
\begin{equation}\mathcal{D}(0,-p)\mathcal{D}(-x,0)\mathcal{D}(0,p)\mathcal{D}(x,0) = e^{i x p } ,\label{glph}\end{equation}
which follows from Eq.~(\ref{comd}), with $xp$ the area enclosed.
\subsection{The qubus computational model}
\begin{figure}[htb!]
 \center
 \includegraphics[width=8cm]{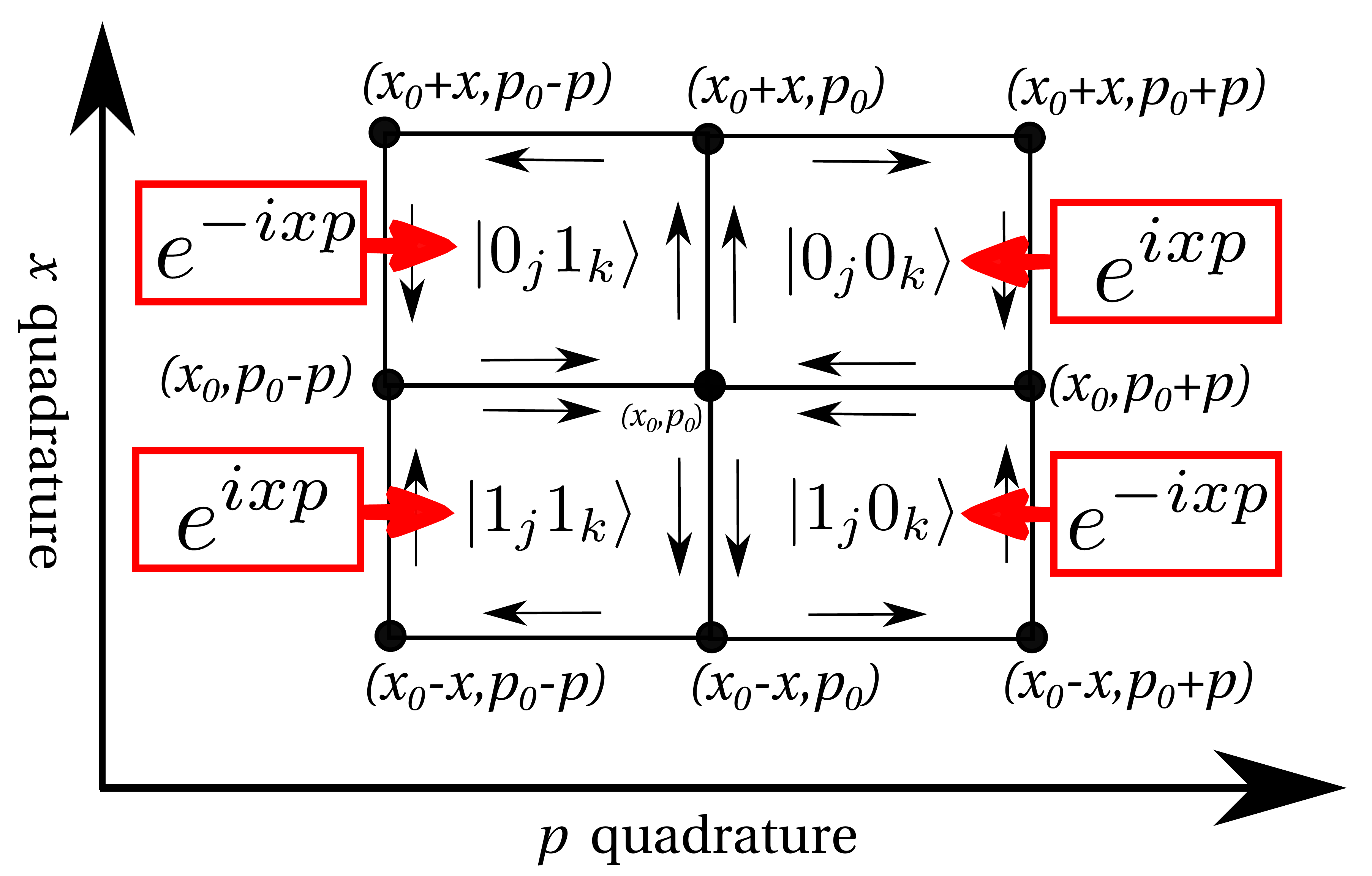}
 \caption{\label{pssqr}(color online) The gate of Eq.~(\ref{rect}) represented in the phase space $\mathbb{R}^2$. The implementation of the gate is independent of the initial state of the field-mode. The phase created is given by $e^{\pm i xp}$ where the phase takes the positive (negative) sign if the loop is traversed clockwise (anti-clockwise). This gate is local equivalent to the controlled phase gate $CR(4xp)$.}
\end{figure}

We now give a brief review of qubus computation based on controlled displacements \cite{spiller2006quantum,brown2011ancilla, louis2007efficiencies}. We take an interaction between a field-mode bus and the $j^{th}$ register qubit of the form
\begin{equation} \mathcal{D}^j(x,p):=C^j(\mathcal{D}(x,p),\mathcal{D}(-x,-p)).\end{equation}
A gate between the register qubits $j$ and $k$ can then be implemented via the ancilla-mediated sequence
\begin{equation}\mathcal{D}^{k}(0,-p)\mathcal{D}^{j}(-x,0)\mathcal{D}^{k}(0,p)\mathcal{D}^{j}(x,0) = e^{i x p  Z_j \otimes Z_k },\label{rect} \end{equation}
which follows directly from Eq.~(\ref{glph}) and is represented pictorially in Fig.~\ref{pssqr}. This two-qubit gate is locally equivalent \cite{makhlin2002nonlocal} to the controlled phase gate $CR(4 xp)$, via local rotations of $ R(-2xp)$ on each computational qubit with the choice of $xp=\pi/4$ giving the maximally entangling gate $CZ$. As any entangling gate in conjunction with a universal set of single-qubit unitaries is universal for quantum computation \cite{brylinski2002universal}, if such a single-qubit gate set can be applied directly to the register \cite{noteA} this is a universal model of AMQC.
\newline
\indent
Using the gate method shown above, $n$ controlled rotation gates can be implemented on a register of qubits using $4n$ bus-qubit interactions - 4 for each gate. However, with certain gate sequences, it is possible to reduce this number by utilising the geometric nature of the gates \cite{louis2007efficiencies,brown2011ancilla,noteE}. For example, $n$ controlled rotations (of arbitrary angle) with one target and $n$ control qubits can be implemented with $2(n+1)$ bus-qubit interactions by first interacting each of the control qubits with the bus via a controlled displacement in one of the quadratures, then interacting the target qubit with the bus via a controlled displacement in the other quadrature with the gate completed by the conjugate of these displacements in sequence  \cite{louis2007efficiencies}. Labelling the control qubits $1 - n$ and the target qubit with the symbol $t$, this is implemented with the interaction sequence
\begin{equation} \mathcal{D}^{t}(0,-p) \cdot  \mathcal{D}^{\text{sq}^{c}_-} \cdot \mathcal{D}^{t}(0,p) \cdot \mathcal{D}^{\text{sq}^c_+}  \\=  \prod_{k=1}^{n} e^{ i \theta_k Z_k \otimes Z_{t}} \label{speedup}, \end{equation}
where $\mathcal{D}^{\text{sq}^c_{\pm}} = \prod_{k=1}^{n} \mathcal{D}^k(\pm x_k,0) $ and $\theta_k=x_kp$. 
By replacing the displacements controlled by the target qubit $t$ with sequences of displacements in the same quadrature controlled by a set of $m$ target qubits we can implement a gate between each of the $m$ target qubits and each of the $n$ control qubits (a total of $m \times n$ gates) using only $2(n+m)$ operations. With the control qubits labelled as before and the target qubits labelled $(n+1) - (n+m)$ we may write this as the interaction sequence
\begin{equation}\mathcal{D}^{\text{sq}^t_-} \cdot  \mathcal{D}^{\text{sq}^c_-} \cdot \mathcal{D}^{\text{sq}^t_+}   \cdot \mathcal{D}^{\text{sq}^c_+} =  \prod_{j=n+1}^{n+m} \prod_{k=1}^{n} e^{ i \theta_{jk} Z_k \otimes Z_j} \label{speedup1}, \end{equation}
where $\mathcal{D}^{\text{sq}^t_{\pm}} = \prod_{j=n+1}^{n+m} \mathcal{D}^j(0,\pm p_j) $ and $\theta_{jk}=x_kp_j$. 
  Using similar techniques, the number of operations required to implement a quantum Fourier transform (QFT)-like structured quantum circuit acting on $n$ qubits can be reduced from a scaling of $n^2$ to a scaling of $n$ \cite{noteE,brown2011ancilla}. We now introduce a computational model based on geometric phases created in the phase space of an ancilla qudit.

\section{Qudit ancilla-mediated quantum computation \label{dqubus}}
\subsection{Phase space formalism} 
\begin{figure}[htb!]
 \center
 \includegraphics[width=9cm]{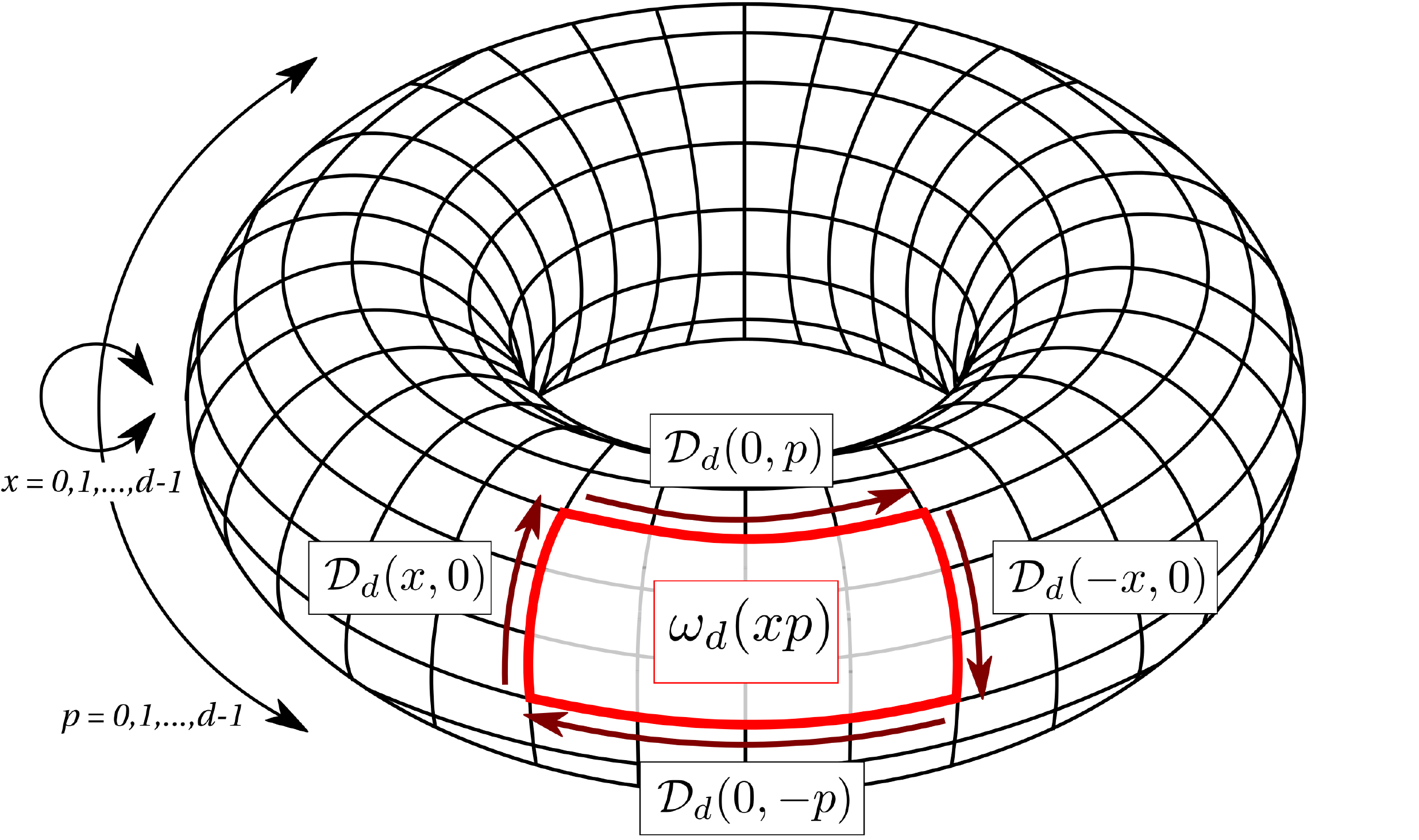}
 \caption{\label{quditphasespace}(color online) The toroidal phase space of a $d$ dimensional qudit: $\mathbb{Z}(d) \times \mathbb{Z}(d)$. In analogy with a field-mode in the phase space $\mathbb{R} \times \mathbb{R}$, a phase is created via displacements around a closed loop. }
\end{figure}

We first consider the phase space and displacement operator for a qudit, a system with a $d$-dimensional Hilbert space, $\mathcal{H}_d$. The  \emph{generalised Pauli operators} for a qudit, denoted $Z_d$ and $X_d$, obey the relation
\begin{equation}Z_d^{p}X_d^x=\omega_d(xp) X_d^{x}Z_d^p, \label{dweyl}\end{equation}
where $x,p \in \mathbb{Z}$  \cite{wootters1987wigner,vourdas2004quantum}.
Take $\ket{m}_x$ and $\ket{m}_p$ with $m \in \mathbb{Z} (d)$ to be two orthonormal bases of  $\mathcal{H}_d$ related by a Fourier transform, i.e.  $\ket{m}_p := F \ket{m}_x$ where $F$ is given by
 \begin{equation} F: = \frac{1}{\sqrt{d}} \sum_{m,n} \omega_d(mn) \ket{m}\bra{n}_x.\end{equation}
The generalised Pauli operators can then be defined as
\begin{equation} X_d := \exp \left(-i \frac{2 \pi}{d} \hat{p}_d \right), \hspace{0.6cm} Z_d := \exp \left(i \frac{2 \pi}{d} \hat{x}_d \right),\end{equation}
where $\hat{x}_d$ and $\hat{p}_d$ are `position' and `momentum' operators given by 
\begin{equation} \label{xp} \hat{x}_d := \sum_{m=0}^{d-1} m \ket{m}\bra{m}_x, \hspace{0.6cm} \hat{p}_d := \sum_{m=0}^{d-1} m \ket{m}\bra{m}_p. \end{equation}
The phase space defined by these operators and bases is the toroidal periodic $\mathbb{Z}(d) \times \mathbb{Z}(d)$ lattice, a torus with $d^2$ discrete points. The operators $X_d^x$  and $Z_d^p$ create translations in position and momentum by $x$ and $p$ discrete lattice points respectively and they are periodic, i.e. $X_d^d=Z_d^d=\mathbb{I}$ \cite{vourdas2004quantum}.  
A displacement operator on this phase space can be defined by \cite{noteB,klimov2009discrete}
\begin{equation} \mathcal{D}_d(x,p) :=\omega_d(-2^{-1} xp)  Z_d^{p} X_d^{x}, \label{Ddisp1}\end{equation}
 where $x,p \in \mathbb{Z}$. Furthermore, it obeys
\begin{equation} \mathcal{D}_d(x_1,p_1)\mathcal{D}_d(x_2,p_2) = \omega_d(\phi)\mathcal{D}_d(x_1+x_2,p_1+p_2), \label{combd2} \end{equation}
where $\phi= 2^{-1}(  x_1 p_2 -p_1 x_2)$.
If we implement displacements around a closed loop in this phase space a phase is created, in particular orthogonal displacements give
\begin{equation}\mathcal{D}_d(0,-p)\mathcal{D}_d(-x,0)\mathcal{D}_d(0,p)\mathcal{D}_d(x,0) = \omega_d(xp), \label{squaredis1} \end{equation}
which is represented graphically on the torus $\mathbb{Z}(d) \times \mathbb{Z}(d)$ in Fig.~\ref{quditphasespace}. If we consider the phase space points in each direction to be separated by a distance of $\sqrt{ 2\pi/d}$, the phase created is then $e^{ \pm i \mathcal{A}}$ where $\mathcal{A}$ is the area enclosed in phase space and the sign depends on the direction that the path is traversed. Hence, the phases that can be created are the $d$ integer powers of $\omega_d$. 
\newline
\indent
A generalisation of $Z_d$ to arbitrary rotations can be obtained by taking
\begin{equation} R_d(\theta) := \exp \left( i \theta \hat{x}_d \right)= \sum_n e^{i n\theta } \ket{n}\bra{n}_x, \label{Zdt}\end{equation}
where $\theta \in \mathbb{R}$. Clearly $Z_d=R_d(2 \pi / d)$. 
We then have that
\begin{equation} \mathcal{D}_d(-x,0) R_d(\theta) \mathcal{D}_d(x,0)\ket{m}_x =  e^{i\theta (x+m)_d  }\ket{m}_x, \label{sq} \end{equation}
for $x \in \mathbb{Z}$, $m \in \mathbb{Z}(d)$ and where the subscript $d$ denotes that the summation is modulo $d$.
Hence as $\theta \in \mathbb{R}$ any phase $\phi \in \mathbb{R}$ may be created by picking a suitable initial qudit state, position displacement and rotation angle $\theta$, for example (independent of the dimension of the qudit) we may take $m=0$, $x=1$ and $\theta=\phi$. A final $R_d(-\theta)$ operator may be included which would implement a phase of $e^{-im\theta}$ but this is no longer necessary due to the operator acting on a specific initial ancilla state (in contrast to the initial state independent Eq.~(\ref{squaredis1})).
 
\subsection{The computational model \label{compd}} 
We can implement a model of universal ancilla-mediated quantum computation by introducing an interaction between an ancilla qudit and the $j^{th}$ computational qubit of the form
\begin{equation} \mathcal{D}^j_{d} (x,p):= C^j\mathcal{D}_{d}(x,p), \label{ind} \end{equation}
which has analogous properties to the ancilla-register interaction in the qubus model \cite{NoteD}.
From Eq.~(\ref{sq}), we have that
\begin{equation} \mathcal{D}_{d}^{k}(0,-p) \mathcal{D}_{d}^{j}(-x,0) \mathcal{D}_{d}^{k}(0,p) \mathcal{D}_{d}^{j}(x,0)= C^j_kR ( \theta ) ,\label{rect2}\end{equation}
where $\theta=  2\pi xp / d$. As this is a two-qubit entangling gate as long as $xp \neq n d$ for any integer $n$ this is universal for quantum computation on the register with the addition of single-qubit gates on the register.
\newline
\indent
It has already been shown that there are computational advantages that can be gained from using ancillary qudits to aid a computational model \cite{ionicioiu2009generalized,ionicioiu2008generalized,ralph2007efficient}. We consider the generalised Toffoli gate which maps the basis states of $n$ control and one target qubit to
\begin{equation} \ket{q_1,q_2...,q_n}_n \ket{\varphi}_t\rightarrow  \ket{q_1,q_2...,q_n}_n U^{q_1 \cdot q_2 \cdot ... \cdot q_n} \ket{\varphi}_t,\end{equation}
for some $U \in U(2)$, where  $q_j=0,1$ denotes the state of the $j^{th}$ qubit and $\ket{\varphi}_t$ is the state of the target qubit. In particular, it has been shown that generalised Toffoli gates can be implemented by only two interactions between each control qubit and the ancillary qudit if a gate controlled on the state of the qudit may also be implemented \cite{ionicioiu2009generalized}. Using the formalism of controlled displacements, and labelling the control qubits $1 - n$, the target qubit $t$ and denoting the initial and final state of the register by $\ket{\psi_i}$ and $\ket{\psi_f}$, this can be achieved using a sequence of the form
\begin{equation}  \mathcal{D}_d^{\text{sq}^c_-}  \cdot \boldsymbol{C}^{n}_tU  \cdot \mathcal{D}_d^{\text{sq}_+^{c}}   \ket{\psi_i} \ket{0}_x =  \ket{\psi_f} \ket{0}_x\end{equation}
where $ \mathcal{D}_d^{\text{sq}^c_{\pm}}= \prod_{k=1}^{n} \mathcal{D}_d^k(\pm x_k,0)$ and $ \boldsymbol{C}^{n}_tU$ is a gate that applies $U$ to the target qubit $t$ if the ancilla is in the state $\ket{n_d}_x$ (again the subscript denotes modulo $d$). If $x_k=1$ for all $k$ and $d > n$ then this applies a generalised Toffoli gate to the register. This utilises the ability of controlled displacements to encode information about the number of register qubits in the state $\ket{1}$ into the \emph{orthogonal} basis states of the qudit. This orthogonality then facilitates gates controlled on this global property of the register qubits. This is in contrast to the use of the continuous variable nature of the field-mode in the computational model of Section \ref{fmqubus}. 
\newline
\indent
The reductions in the number of operations required for certain gate sequences in the qubus model rely on the geometric nature of the phases used for the gates. We have seen that we may also consider the phases created by displacements of the form $\mathcal{D}_d(x,p)$ around closed loops in the finite and periodic lattice phase space of a qudit to also be geometric (in a certain sense \cite{noteC}) and hence we will show that similar computational savings are possible in this model. We have seen that the geometric phases that can be created from displacements of the form $\mathcal{D}_d(x,p)$ are the $d$ integer powers of $\omega_d$. Hence, if a gate sequence is composed only of controlled rotations of the form $CR(2 n \pi /d)$ for integer $n$ a $d$-level qudit can also implement this gate sequence with the same number of operations as in the qubus model (ignoring the additional local corrections required in the qubus model). We illustrate this with a sequence, analogous to that in Eq.~(\ref{speedup}), in which $n$ controlled rotations with one target qubit and $n$ control qubits are implement with $2(n+1)$ operations. With the target qubit again labelled $t$ and the control qubits labelled $1 - n$ we have that
\begin{equation}\mathcal{D}_d^{t}(0,-p) \cdot \mathcal{D}_d^{\text{sq}^c_-} \cdot \mathcal{D}_d^{t}(0,p) \cdot \mathcal{D}_d^{\text{sq}^c_+}  =  \prod_{k=1}^{n} C^k_t R ( \theta_k ) \label{speedupd}, \end{equation}
where $\mathcal{D}_d^{\text{sq}^c_{\pm}}$ is given earlier and $\theta_k=2 \pi x_kp/d$.  Similarly, the sequence of Eq~(\ref{speedup1}), in which $m \times n$ controlled rotation gates can be implemented in $2(m+n)$ operations, is also applicable to this model. If we label the control qubits as before and the target qubits $(n+1) - (n+m)$ we may write this as the interaction sequence
\begin{equation} \mathcal{D}_d^{\text{sq}^t_-} \cdot  \mathcal{D}_d^{\text{sq}^c_-} \cdot \mathcal{D}_d^{\text{sq}^t_+} \cdot \mathcal{D}_d^{\text{sq}^c_+} =  \prod_{j=n+1}^{n+m} \prod_{k=1}^{n} C^k_jR( \theta_{jk}) , \end{equation}
where $\mathcal{D}_d^{\text{sq}^t_{\pm}} = \prod_{j=n+1}^{n+m} \mathcal{D}_d^j(0,\pm p_j) $ and $\theta_{jk}=2 \pi x_kp_j / d$. 
\newline
\indent
In some gate sequences only controlled rotations that are maximally entangling, and hence locally equivalent to $CZ$, are present. If this is the case a qubit ancilla, i.e. $d=2$, is sufficient to implement any of the sequences of the qubus model.
\newline
\indent
We have restricted the analysis here to a model with ancilla-register interactions that are controlled displacements operators and hence is directly analogous to the qubus model. In Appendix \ref{ap2} we discuss a model in which $D^j(0,p)$ gates are generalised to controlled $R_d(\theta)$ interactions. In this case we can create controlled rotation gates with arbitrary phase (rather than only integer powers of $\omega_d$)  between pairs of qubits by using the equality of Eq.~(\ref{sq}) (and ancilla preparation). However we show this does not imply the qubus decomposition results hold when the required phases are not integer powers of $\omega_d$ (although alternative interesting multi-qubit gates can be implemented efficiently). Therefore, the dimensionality of the qudit, although not relevant to the universality of the model, affects the power of the model to reduce the number of ancilla-register interactions required to implement certain gate sequences.

\subsection{Implementation}
An ancilla-register interaction $C^j(R_d(\theta),R_d(-\theta))$ can be generated, up to irrelevant phase factors, by applying the Hamiltonian
\begin{equation} H_{d} =  Z_j \otimes S_z, \label{Hd}\end{equation}
for a time $t=\theta$, where $S_z$ is the effective $z$-spin operator for a $d$ level qudit given by
\begin{equation} S_z = \mbox{diag} (s,s-1,...,-s+1,-s), \end{equation}
where $s=(d-1)/2$ and `diag' is the diagonal matrix in the position basis. By acting the local unitary $R_d(-\theta)$ on the ancilla and taking appropriate values for $\theta$ we may implement any $\mathcal{D}^j_{d}(0,p)$. As $X_d=F^{\dagger} Z_d F$ \cite{vourdas2004quantum}, we have that
\begin{equation} \mathcal{D}_d^{j}(x,0)  = F^{\dagger} \cdot \mathcal{D}_d^{j}\left(0, x \right)\cdot F, \end{equation}
and hence displacements in both quadratures can be implemented via the interaction of Eq.~(\ref{Hd}) and local operations on the qudit.
\newline
\indent
Physical systems that are used as qubits are often restrictions of higher dimensional systems to a 2-dimensional subspace and hence many of these systems are naturally suited to a $d$-level qudit structure  \cite{devitt2007subspace}. Qudits have been demonstrated in various physical systems, including superconducting \cite{neeley2009emulation}, atomic \cite{mischuck2012control} and photonic systems, where in the latter the qudit is encoded in the linear  \cite{lima2011experimental,rossi2009multipath} or orbital angular momentum \cite{dada2011experimental} of a single photon. A further possible realisation of a qudit is in the Fock states of a field mode which can be coupled to individual qubits via the Jaynes-Cummings model  \cite{mischuck2013qudit}. The dispersive limit of the Jaynes-Cummings model results in an effective coupling of the form
\begin{equation} H_{\text{eff}} = Z_j \otimes a^{\dagger}a,     \end{equation}
which with this qudit encoding is equivalent to the Hamiltonian $H_{d}$. Furthermore it has been suggested \cite{ionicioiu2009generalized} that controlled $Z_d$ gates may be realisable in the dispersive limit of the generalised Jaynes-Cummings model, which describes the coupling of a spin-$s$ particle to a field mode, with a photonic qubit encoded in the field mode.
\newline
\indent
An alternative candidate physical system is an ensemble of $N$ qubits on which we may define the collective spin operators $J_{\mu}= \sum_{j=1}^N \mu_j$ with $\mu=X,Y,Z$ and $J^2=J_x^2+J_y^2+J_z^2$ which obey the $SU(2)$ commutation relations. The simultaneous eigenstates of $J_z$ and $J^2$ are known as the Dicke states and when such an ensemble is restricted to the $N+1$ dimensional subspace which is symmetric with respect to qubit exchange it may be considered to be a $d=N+1$ dimensional qudit with a basis given by the symmetric Dicke states of the ensemble. Indeed, there have been proposals for qubit ensembles to be coupled to computational qubits in the context of utilising the collective ensemble states as a quantum memory \cite{rabl2006hybrid,marcos2010coupling,lu2013quantum,petrosyan2009reversible}. A particularly promising candidate for such a ensemble-qubit hybrid system is in the coupling of an NV center ensemble in diamond to a flux qubit with coherent coupling between such systems having been experimentally demonstrated \cite{zhu2011coherent} and we will return to this later.  An alternative formalism for the $N+1$ dimensional symmetric subspace of such an ensemble is to consider the $SU(2)$ or spin coherent states, and in the next section we show that with a suitably defined displacement operator on these states that we can implement a continuous-variable based spin-ensemble ancilla-mediated model. 

\section{Spin coherent state ancilla-mediated quantum computation}
\label{spinstates}
\subsection{Phase space formalism}
\begin{figure}[htb!]
 \includegraphics[width=5cm]{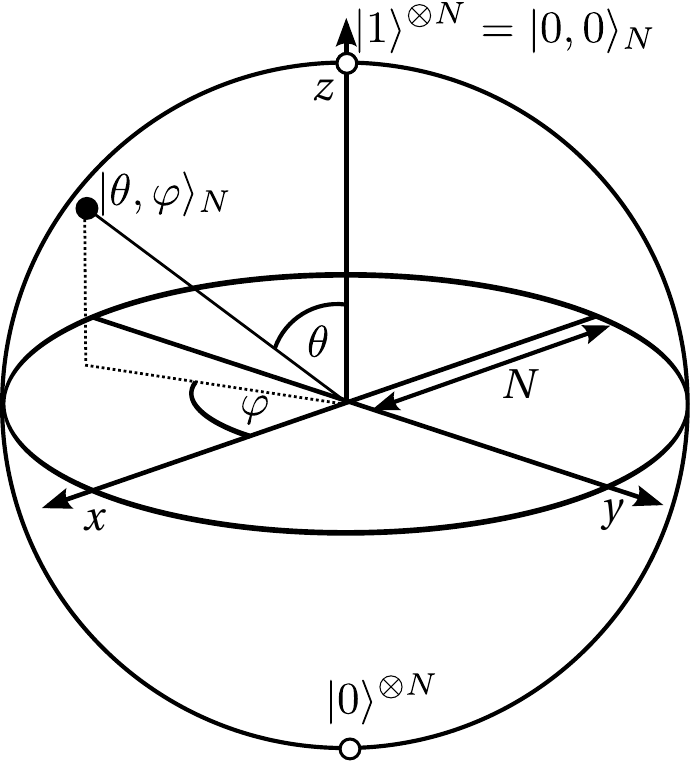}
 \caption{ \label{blochspherea} The spin coherent states of $N$ qubits can be represented on a Bloch sphere of radius $N$. The displacement operator of Eq.~(\ref{scdisp}) can be interpreted as a rotation around some vector in the $xy$-plane. The north pole represents the reference state $\ket{0,0}_N$ and can be considered to be the phase space origin. }
\end{figure}

\begin{figure}[htb!]
 \includegraphics[width=5.1cm]{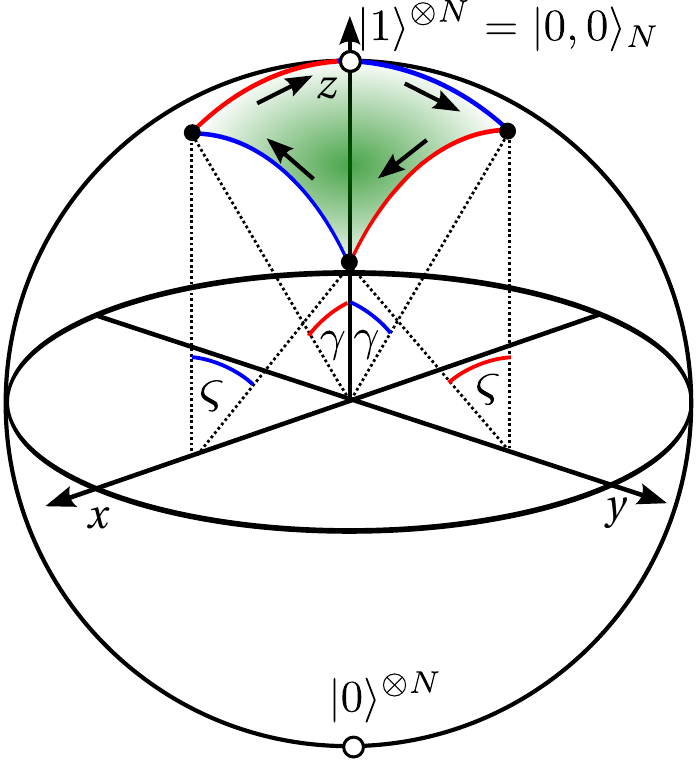}
 \caption{ \label{blochsphereb}(color online) A geometric phase is obtained by the displacement of a spin coherent state around a closed loop. Here we depict orthogonal displacements which correspond to rotations around the $x$ and $y$ axes. We can relate the rotation angles $\gamma$ and $\varsigma$ to the variables $\eta$ and $\tau$ from Eq.~(\ref{scsphase2}) via the stereographic projection.}
\end{figure}

We first introduce the spin coherent states of a collection of $N$ qubits, also referred to as $SU(2)$ or atomic coherent states \cite{radcliffe1971some,arecchi1972atomic,gazeau2009coherent}. We define a displacement (or rotation \cite{arecchi1972atomic,gazeau2009coherent}) operator by
\begin{equation} \mathcal{D}_N(\theta, \varphi) := e^{i\left( \frac{\theta}{2} \sin \varphi J_x - \frac{\theta}{2} \cos \varphi  J_y  \right)} ,\label{scdisp} \end{equation}
where $\theta,\varphi \in \mathbb{R}$ \cite{zhang1990coherent}.  A spin coherent state of $N$ qubits is then defined as
\begin{equation} \ket{\theta,\varphi}_N := \mathcal{D}_N(\theta,\varphi) \ket{0,0}_N, \end{equation}
where the reference state is taken to be $\ket{0,0}_N=\ket{1}^{\otimes N}$. A spin coherent state is a separable state of $N$ qubits in the same pure state \cite{dooley2013collapse} which may be written as
\begin{equation}\ket{\theta, \varphi}_{N} = \left( \cos \frac{\theta}{2} \ket{1} - e^{-i \varphi} \sin \frac{\theta}{2} \ket{0} \right)^{\otimes N},\end{equation}
or alternatively they may be expressed in terms of the symmetric Dicke states that are a basis for the $N+1$ dimensional symmetric subspace. 
The phase space of $N$ qubits restricted to such states can be represented on a Bloch sphere of radius $N$, as depicted in Fig.~\ref{blochspherea}, and the displacement operator can be interpreted as a rotation around some vector in the $xy$-plane.
\newline
\indent
We may introduce an alternative parameterisation for the spin coherent states, analogous to writing a field-mode coherent state in terms of a complex number $\alpha$, that is a stereographic projection of the sphere onto the complex plane.
We take  $\zeta=-e^{-i \varphi} \tan \frac{\theta}{2}$ \cite{gazeau2009coherent} with which the spin coherent states can be expressed as
\begin{equation} \ket{\zeta}_N = \left( \frac{ \ket{1} + \zeta \ket{0} }{\sqrt{1+|\zeta|^2}} \right) ^{\otimes N}. \end{equation} 
In this parameterisation the displacement operator becomes
\begin{equation} \mathcal{D}_N(\zeta) = \left( \frac{I_2 + \zeta \sigma_+- \zeta^* \sigma_- } {\sqrt{1+|\zeta|^2}} \right) ^{\otimes N}, \end{equation}
where $\sigma_{\pm}=\frac{1}{2}(X \pm i Y)$. It is straight forward to confirm that $\mathcal{D}_N(\zeta) \ket{0}_N = \ket{ \zeta}_N$ and furthermore we have the identity
\begin{equation} \mathcal{D}_N(\zeta_2)\mathcal{D}_N(\zeta_1) \ket{0}_N = e^{i N \phi(\zeta_1,\zeta_2)} \left| \frac{ \zeta_1+\zeta_2}{1- \zeta_1\zeta_2^*} \right \rangle_N, \label{comscs}\end{equation}
where we have that
\begin{equation}\phi(\zeta_1,\zeta_2) =  \frac{1- \zeta_1 \zeta_2^*} { |1- \zeta_1\zeta_2^*| }.  \end{equation}
\newline
\indent
As in the the case of a field-mode or qudit, closed loops in phase space create geometric phases. Displacements around the orthogonal $x$ and $y$ axes are given by taking $\cos\varphi =0$ ($\zeta \in \mathbb{R}$) and $\sin \varphi=0$ ($ \zeta \in \mathbb{I}$) respectively. We consider a sequence of orthogonal displacements, acting on a coherent state, of the form
\begin{equation} \mathcal{D}_N(-i\zeta_4)\mathcal{D}_N(-\zeta_3)\mathcal{D}_N(i\zeta_2)\mathcal{D}_N(\zeta_1) \ket{0}_N = e^{i \phi_t} \ket{\zeta_t}_N, \label{scsphase} \end{equation}
where $\zeta_j \in \mathbb{R}$, $j=1-4$. In order to create a geometric phase and no overall displacement we require that $\zeta_t=0$. If the phase space geometry is flat, as in the case of a field-mode, we can take $\zeta_1=\zeta_2=\zeta_3=\zeta_4$ (as in Eq.~(\ref{glph}) with $x=p$). However, on the surface of a sphere this is not the case and if we restrict $\zeta_j$ such that $\zeta_4=\zeta_1=\eta$ then, using Eq.~(\ref{comscs}), it can be shown that to satisfy $\zeta_t=0$ we must take $\zeta_2=\zeta_3=\tau(\eta)$, where
\begin{equation} \tau(\eta) = \frac{1-\eta^2-\sqrt{\eta^4-6\eta^2+1}}{2 \eta}.  \label{scsphase2} \end{equation}
The corresponding phase, $\phi_t$, is given by
\begin{equation}  \phi_t =N \tan^{-1} \left( \frac{2\eta \tau +\tau^2-\eta^2}{1+2\eta\tau-\eta^2\tau^2} \right) .\label{scsphase3} \end{equation}
That there does exist such a $\tau$ and that $\tau \neq \eta$ can be seen schematically from Fig.~\ref{blochsphereb}. 
\subsection{The Computational model}
\begin{figure}[htb!]
    \includegraphics[width=8cm]{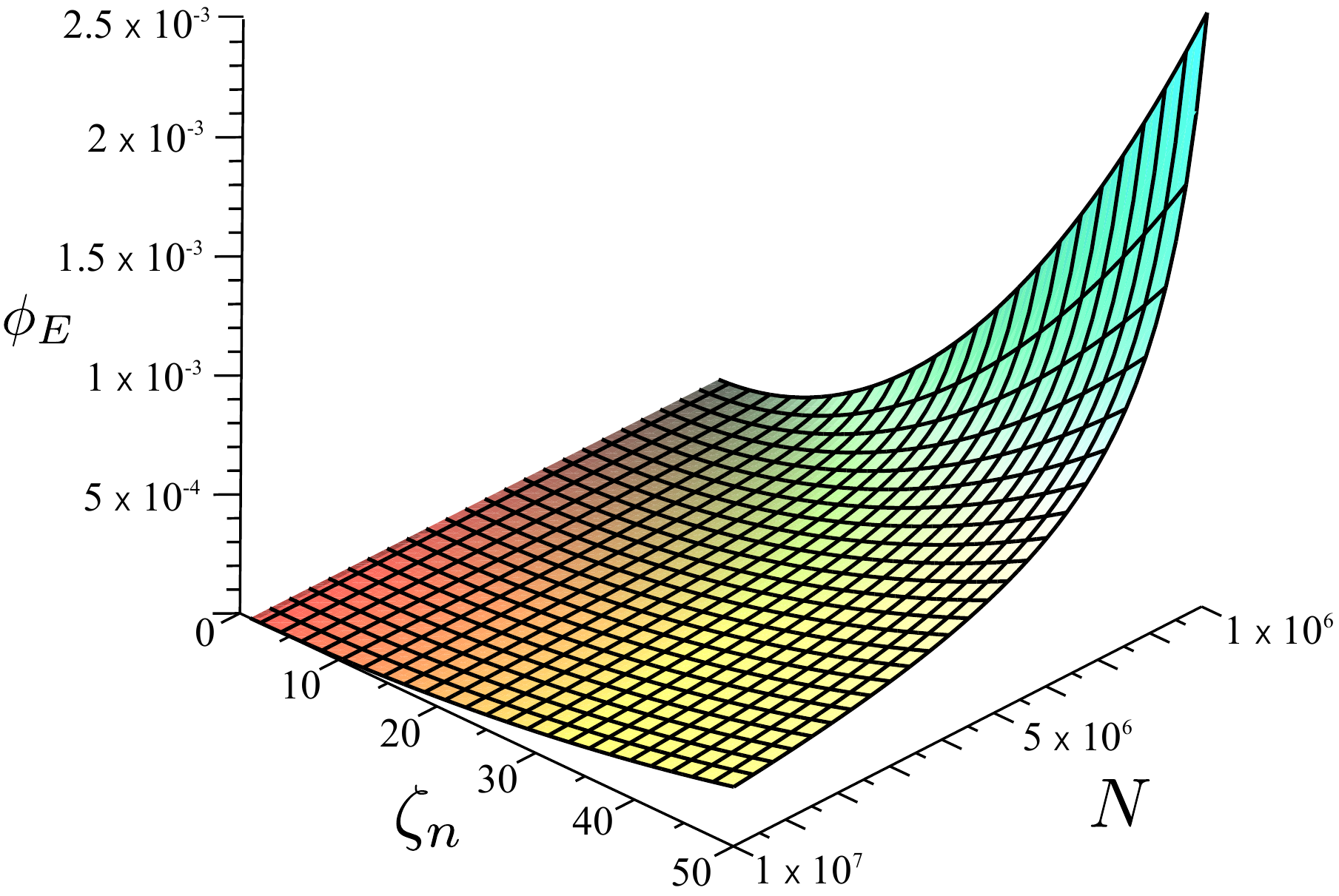}
   \caption{\label{phaseerror}(color online) The fractional error in the phase, $\phi_{E} = \frac{\zeta_n^2 - \phi_f}{\zeta_n^2}$, as a function of $\zeta_n$ and $N$ where $\phi_f$ given by Eq.~(\ref{endphi}). The range of $N$ includes realistic values for the coherent manipulation of spin-ensembles  \cite{zhu2011coherent,christensen2013quantum}. }
\end{figure}

\begin{figure}[htb!]
   \includegraphics[width=8cm]{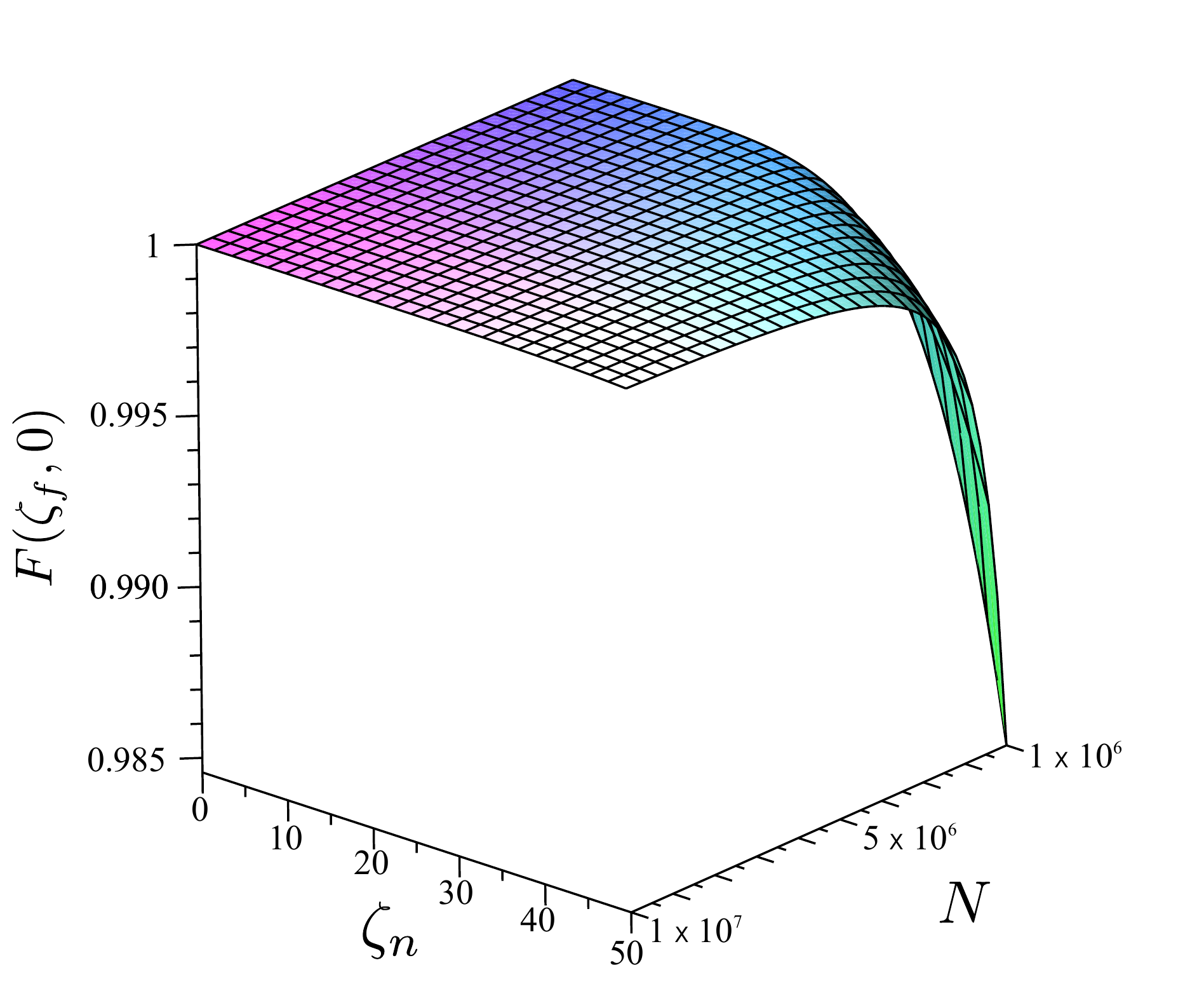}
     \caption{(color online) A plot of the the fidelity $F(\zeta_f,0)$, given in Eq.~(\ref{fidzeta}) as a function of $\zeta_n$ and $N$. The range of $N$ includes realistic values for the coherent manipulation of spin-ensembles  \cite{zhu2011coherent,christensen2013quantum}. \label{fidelity}}
\end{figure}

We now show how these geometric phases may be used to implement a model of ancilla-mediated quantum computation. We consider an ancilla spin-ensemble and introduce an interaction between this ancilla and the $j^{th}$ register qubit of the form
\begin{equation} \mathcal{D}^j_N(\zeta) := C^j (\mathcal{D}_N(\zeta),\mathcal{D}_N(-\zeta)) .\end{equation}
A computational gate between a pair of register qubits $j$ and $k$ can be implemented using the interaction sequence
\begin{equation} \mathcal{D}_N^{k}(-i\eta)\mathcal{D}_N^{j}(-\tau)\mathcal{D}_N^{k}(i\tau)\mathcal{D}_N^{j}(\eta) \ket{\psi}  \ket{0}_N \\=  \ket{\varphi} \ket{0}_N,\label{rect3}\end{equation}
where $\ket{\psi}$ is the initial state of the qubits $j$ and $k$, $\ket{\varphi} =\exp(i \phi_t Z \otimes Z) \ket{\psi}$ and $\tau$ and $\phi_t$ are given by Eq.~(\ref{scsphase2}) and Eq.~(\ref{scsphase3}) respectively. As the gate implemented on the register is identical to that in the qubus model, ancilla-register interactions of this form can implement a universal ancilla-mediated model of quantum computation with the addition of single-qubit gates on the register.
\newline
\indent
In section \ref{fmqubus} we reviewed the methods that may be used in qubus computation to reduce the number of bus-qubit interactions required in certain gate sequences from the upper limit of $4n$ for $n$ controlled rotations. The schemes to reduce the number of operations required for a particular gate sequence, such as those given in Eq.~(\ref{speedup}) and Eq. (\ref{speedup1}), require that more than two register qubits are entangled with the bus at the same time, and in particular more than one qubit is entangled with each quadrature of the bus. In order to create a closed phase space path via displacements on a spin coherent state, it is necessary to take into account the curvature of the phase space as quantified by Eq.~(\ref{scsphase2}). However, this is not possible when there are multiple qubits entangled with either quadrature as then different parts of the spin coherent state superposition are different distances from the phase space origin (the north pole). Hence, not all the phase space paths can be perfectly closed and the ancilla will remain entangled with the register qubits if such sequences of controlled displacements are applied.
\newline
\indent
 In the limit that $N \to \infty$, a spin coherent state is equivalent to a field mode  \cite{dooley2013collapse}. In particular, we show in Appendix~\ref{sclimit} that
\begin{equation} \lim_{N \to \infty}\mathcal{D}_N\left(\frac{ \zeta}{\sqrt{2N}} \right)   = \mathcal{D}(\Re(\zeta),\Im(\zeta)) , \label{limdisp}\end{equation}
where $\Re(\zeta)$ and $\Im(\zeta)$ denote the real and imaginary parts of $\zeta$ respectively and $\mathcal{D}(x,p)$ is the field-mode displacement operator of Eq.~(\ref{displace}). Hence, in this limit all the gate sequences of qubus computation can be implemented with a spin-ensemble ancilla. Although for finite $N$ these sequences will not create the exact gates required, and the ancilla will remain partially entangled with the register, they will implement the desired gates with some fidelity that tends to unity as $N \to \infty$.
\newline
\indent
We now consider the intrinsic error in a gate sequence given that $N$ is finite. We do this by considering the error accumulated when we do not take account of the curvature of the phase space and treat the ancilla as a field-mode. We initially consider the specific example of implementing $m \times n$ controlled rotations between each of $n$ control and $m$ target qubits using $2(n+m)$ ancilla-register interactions. We do this by taking the operator sequence of Eq.~(\ref{speedup1}) and letting $\mathcal{D}^j(x,p) \to \mathcal{D}_N^j\left((x+ip)/ \sqrt{2N}\right)$.  For simplicity we take $m=n$ and act this sequence on an initial state $\ket{\psi}\ket{0}_N$ giving some resultant state of the whole system, $\ket{\psi_f}_{G}$, where
\begin{equation}\ket{\psi_f}_{G}= \mathcal{D}_N^{\text{sq}^t_{-}} \cdot \mathcal{D}_N^{\text{sq}^c_-} \cdot \mathcal{D}_N^{\text{sq}^t_+} \cdot \mathcal{D}_N^{\text{sq}^c_+} \ket{\psi} \ket{0}_N  \label{speedupscs}, \end{equation}
in which $\mathcal{D}_N^{\text{sq}^c_{\pm}} = \prod_{k=1}^{n} \mathcal{D}^k_N(\pm x_k /\sqrt{2N}) $ and $\mathcal{D}_N^{\text{sq}^t_{\pm}} = \prod_{k=n+1}^{2n} \mathcal{D}^k_N(\pm ip_k/ \sqrt{2 N}) $. Given that in the limit $N \to \infty$ this sequence is equivalent to Eq.~(\ref{speedup1}) with $m=n$, we wish to estimate how well $\ket{\psi_f}_{G}$ approximates $\hat{O}\ket{\psi} \ket{0}_N$ where $\hat{O}$ is the operator on the right hand side of Eq.~(\ref{speedup1}) given by
 \begin{equation}  \hat{O}= \prod_{j=n+1}^{2n} \prod_{k=1}^{n} e^{ i \theta_{jk} Z_k \otimes Z_j} ,\end{equation}
  where $\theta_{jk} = x_k p_j$. We consider the conditions under which the errors in the phase and final ancilla state associated with each register basis state are negligible, and hence $\hat{O}$ is well approximated. These phase and ancilla state errors for each register computational basis states will be bounded by the error in the ancilla state that is displaced furthest from the origin. If we fix all $x_k >0$ and all $p_j >0$, the ancilla state displaced furthest from the origin is any of the four ancilla states associated with a basis state in which all the control qubits are all in the same state and similarly all the target qubits are all in the same state. We consider the ancilla state associated with all the register qubits being in the state $\ket{0}$. For simplicity we choose the displacements such that $\sum^{n}_{k=1} x_k=\sum^{2n}_{k=1+n} p_k=:\zeta_n$
  The final state of the ancilla mode associated with this state is given by
  \begin{equation} e^{i \phi_f} \ket{\zeta_f}_N = \mathcal{D}_N(-i\zeta_N)\mathcal{D}_N(-\zeta_N)\mathcal{D}_N(i\zeta_N)\mathcal{D}_N(\zeta_N) \ket{0}_N,  \label{scserrors1}\end{equation}
where $\zeta_N=\zeta_n /\sqrt{2N}$. The final state in the field mode case, and hence the state we wish to approximate, is given by $\phi_f=\zeta^2_n$ and $\ket{\zeta_f}=\ket{0}$ from Eq.~(\ref{glph}) and Eq.~(\ref{limdisp}). Using Eq.~(\ref{comscs}) we can calculate the phase $\phi_f$ and the parameter $\zeta_f$. We have that
  \begin{equation}\phi_f = N \tan^{-1} \left(  \frac{2\zeta_N^2}{1+2\zeta_N^2-\zeta_N^4}\right), \label{endphi} \end{equation}
which we may expand to first order in $1/N$, giving
\begin{equation}\phi_f = \zeta^2_n - \frac{\zeta^4_n}{N} +  \mathcal{O} \left(\frac{1}{N^2} \right).\label{endzeta}  \end{equation}
Hence, for large $N$ we have that $\phi_f \approx \zeta_n^2$ with an error of order $ \frac{1}{N}$ which is negligable when $\zeta_n^4 << N$. In Fig.~(\ref{phaseerror}) we plot the fractional error in the phase, $\phi_E = \frac{\zeta_n^2- \phi_f}{\zeta_n^2}$, as a function of $\zeta_n$ and $N$. From the definition of $\zeta_n$ we see that the size of $\zeta_n$ is related to the number of qubits that can be entangled with this sequence. If we wish to implement a maximally entangling gate between each of $n$ control and $n$ target qubits we have that $\zeta_n = \frac{\sqrt{\pi}}{2} n \approx n$. We see from Fig.~(\ref{phaseerror}) that with $N=10^7$, which has been achieved with the coherent manipulation of NV center ensembles  \cite{zhu2011coherent}, gates between a large number of qubits may be implemented with a low phase error. For example with $\zeta_n=40$ (hence $40^2=1600$ maximally entangling gates can be implemented between 40 control and 40 target qubits using only 160 operations) and $N=10^7$ we have that $\phi_E\approx 2 \times 10^{-4}$.
\newline
\indent
The other intrinsic source of error is due to the ancilla mode not exactly returning to its initial state and remaining entangled with the register. The fidelity between the desired final state, $\ket{0}_N$, and the actual final state $\ket{\zeta_f}_N$, $F(\zeta_f,0)= |\langle \zeta_f,0 \rangle |^2$, is given by
\begin{equation}F(\zeta_f,0)= \left(1+ \frac{8 \zeta_N^6}{(1+\zeta_N^2)^4}\right)^{-N}, \label{fidzeta}\end{equation}
which we may expand to second order in $1/N$, giving
\begin{equation} F(\zeta_f,0)= 1 - \frac{\zeta^6_n}{N^2} + \mathcal{O} \left(\frac{1}{N^3} \right). \end{equation}
Hence, for larger $N$ we have that $ F(\zeta_f,0) \approx 1$ with an error of order $1/N^2$. This fidelity is shown as a function of $\zeta_n$ and $N$ in Fig.~(\ref{fidelity}). Again, we see that for realistic numbers of spins in the ancillary ensemble the fidelity is very close to unity. Using the same example as above, when $\zeta_n=40$ and $N=10^7$ we have that $1-F(\zeta_f,0) \approx 4 \times 10^{-5}$.
\newline
\indent
In any finite sequence of controlled displacement the state of the ancillary mode will be bounded within some phase space square centred on the origin. Hence, when the errors accrued from traversing this bounding square are negligible, which can be assessed using the above error analysis, the intrinsic errors due to the phase space curvature in such a sequence will also be small.

\subsection{Implementation}
The ancilla-register interaction $\mathcal{D}^j_N(\zeta)$ can be generated by the Hamiltonian
\begin{equation} H_{N}=  Z_j\otimes  X (\phi ), \label{scsham} \end{equation}
where $X(\phi)=\sin \phi J_x + \cos \phi J_y$. Only one value of the parameter $\phi$ is required to create displacements in both quadratures as $U^{\dagger}\cdot e^{i \theta J_x}  \cdot  U=e^{i \theta J_y}$ with $U=(R(\pi/2) H)^{\otimes N}$. As we have already mentioned, a particular promising hybrid system in which to realise an ensemble-qubit coupling is with an ensemble of NV centers coupled to a superconducting flux qubit, such as in the proposals of \cite{marcos2010coupling,lu2013quantum}. Such a coupling has been experimentally realised \cite{zhu2011coherent} with a coupling term of the form 
\begin{equation}H_{\text{coupling}} =Z \otimes J_x + \sum_{k=1}^N \delta_k Z \otimes X_k \end{equation} 
where in this context the $\delta_k$ terms can be considered to be error terms due to the coupling strength varying over the ensemble. A physical setup of the type demonstrated in \cite{zhu2011coherent} has the advantage that the NV centers have an energy spectrum that may allow for gap-tunable flux qubits to sequentially interact with the spin-ensemble by bring them into resonance in turn. Such an ensemble realises either a qudit or a spin-coherent state by restricting the ensemble to its symmetric subspace and hence leakage out of this subspace is an important source of errors. Such leakage can be caused by inhomogeneity in the ensemble, for example if all the NV centers do not have an identical energy gap or in the realistic case of $\delta_k \neq  0$ due to the coupling strength varying over the ensemble. An important topic for future work would be to consider the effect on the computational model of the physically relevant errors, such as those outlined above, within the realistic parameter regimes for a specific realisation.

\section{Conclusions}
We have introduced a model of ancilla-mediated quantum computation based on controlled displacements operators acting on an ancillary qudit. These displacement operators can be considered to create geometric phases in a periodic and discrete phase space. We have shown that this model can harness the computational advantages previously demonstrated in the qubus model, whereby the number of ancilla-register interactions required to implement certain gate sequences can be hugely reduced. Furthermore, using the work of Ionicioiu \emph{et al.} \cite{ionicioiu2009generalized} we have seen that in this model generalised Toffoli gates can also be implemented with a large saving in the number of operations required.
\newline
\indent
  An alternative finite-dimensional formalism with analogies to a field-mode is the spin coherent states of a spin-ensemble. We have shown that with an appropriately defined controlled displacement operator, that can be interpreted in terms of controlled rotations on a Bloch sphere, such an ancillary system may also be used to implement a simple universal model of ancilla-mediated quantum computation. For a finite number of spins making up the spin coherent states, the gate decomposition schemes of qubus computation cannot be exactly implemented in this model. However we show that for realistic numbers of spins these intrinsic errors are small and the gate decompositions implement the desired register gates with a high fidelity. An interesting extension could be to consider ancilla-register interactions that employ more general transformations in $SU(n)$ and in particular investigating an interaction based on the displacement operators for $SU(n)$ coherent states \cite{Nemoto2000generalised}.
\newline
\indent
A source of error relevant to computational models of the type presented here is the propagating of correlated errors in the computational register due to many register qubits being simultaneously entangled with the ancillary system. It has been shown that limiting the number of register qubit entangled with the ancilla at one time and refreshing the ancilla after a certain number of gates (with this number dependant on the strength of the various decoherence mechanisms) can mitigate these errors in the qubus model \cite{horsman2011reduce}. Equivalent results will hold in the models presented here and a specific analysis for the physically relevant decoherence model in a proposed realisation would be interesting future work.

\section*{Acknowledgements}
The authors would like to thank Bill Munro for helpful comments on the manuscript. TJP was supported by a university of Leeds Research Scholarship.

\appendix

\section{ \label{ap2}}
It is possible to implement a controlled rotation between a pair of qubits that is of arbitrary angle by using a controlled $R_d(\theta)$ operators in place of the $\mathcal{D}^j(0,p)$ operators in the sequence of Eq.~(\ref{ind}) if the ancilla may be prepared in the `position' basis. This can be achieved with the sequence
\begin{equation*}   \mathcal{D}^j_d (-1,0)  \cdot C^k_a R_d(\theta) \cdot \mathcal{D}^j_d (1,0) \ket{\psi} \ket{0}_x = C^j_k R(\theta) \ket{\psi} \ket{0}_x,  \end{equation*}
where this equality follows from Eq.~(\ref{sq}) and where a final $C^k_a R(-\theta)$ is not required (but could be included) due to the ancilla preparation in a particular state. 
\newline
\indent
We now show why gate sequence decompositions equivalent to those in qubus computation do not always hold when we allow these continuous gates. Consider a sequence of the form,
\begin{equation*}   \mathcal{D}_d^{\text{sq}^c_{-}}\cdot C^k_a R_d(\theta) \cdot  \mathcal{D}_d^{\text{sq}^c_{+}} \ket{\psi} \ket{0}_x = G (\theta) \ket{\psi} \ket{0}_x,  \end{equation*}
where $ \mathcal{D}_d^{\text{sq}^c_{\pm}}= \prod_{k=1}^{n} \mathcal{D}_d^k(\pm x_k,0)$ as in the main text. This is analogous to Eq.~(\ref{speedupd}) and it clearly holds for some gate $G (\theta)$ acting only on the register qubits as $R_d(\theta)$ is diagonal in the position basis. What is the gate $G (\theta)$? Letting all $x_k=1$ (the generalisation is straightforward) and again using Eq.~(\ref{sq}) we can show that it maps
\begin{equation*} \ket{q_1,...,q_n}\ket{q_t}_t \to e^{i \theta (q_1+...+q_n)_d q_t} \ket{q_1,...,q_n}\ket{q_t}_t  ,    \end{equation*}
where the subscript $d$ denotes modulo arithmetic. When $n < d$ then the modulo arithmetic is equivalent to ordinary arithmetic and hence 
\begin{equation*} G (\theta) = \prod_{k=1}^n C^k_t R (\theta), \end{equation*}
as in the qubus model. However, if $n > d$ then this is not the case. The $ \mathcal{D}_d^{\text{sq}^c_{+}}$ sequence can be considered to encode into the ancillary qudit the value of $q_1+...+q_n$ modulo $d$. In the case of $d=2$ this encodes the parity of the $n$ qubits into the ancilla, and hence for general $d$ this can be seen to be a generalisation of parity to modulo $d$ arithmetic. The size of the rotation on the target qubit is then effectively controlled by this global property of the $n$ control qubits. Note that this has a very similar structure to the technique used to implement the generalised Toffoli gate considered in the main text. A straightforward extension is given by allowing multiple target qubits.

\section{ \label{sclimit}}
Here we review the group contraction of $SU(2)$ which gives the $N \to \infty$ limit of the spin coherent states \cite{radcliffe1971some,arecchi1972atomic,gazeau2009coherent,dooley2013collapse} and show that in this limit the displacement operator of Eq.~(\ref{scdisp}) is equivalent to that of a field mode. The bosonic creation and annihilation operators, denoted $a^{\dagger}$ and $a$, can be defined by $a^{\dagger} :=  \frac{1}{\sqrt{2}}( \hat{x}-i\hat{p})$ and  $a :=  \frac{1}{\sqrt{2}}( \hat{x}+i\hat{p})$. The $J$ spin operators obeying $[J_x,J_y]=2 i J_z$ can be related to those of a bosonic mode by the Holstein-Primakoff transformation \cite{holstein1940field}
\begin{equation*}\frac{J_+}{\sqrt{N}} = a^{\dagger} \sqrt{1-\frac{a^{\dagger}a}{2N}} ,\hspace{0.5cm}  \frac{J_-}{\sqrt{N}} = \sqrt{1-\frac{a^{\dagger}a}{2N}} a, \label{lim} \end{equation*}
 with $J_{\pm} := \frac{1}{2}(J_x \pm  i J_y) = \sum_{j=1}^{N} \sigma_{\pm_j} $, and
\begin{equation*} J_z = a^{\dagger}a - N. \end{equation*}
It then follows that
\begin{equation*}\lim_{N \to \infty}\frac{J_x}{\sqrt{2  N}} = \hat{x},\hspace{1.cm}  \lim_{N \to \infty}\frac{J_y}{\sqrt{2  N}}= -\hat{p}. \label{lim} \end{equation*}
We have that $\zeta=-e^{-i\varphi} \tan\frac{\theta}{2}$ and hence from the definition of $D_N(\theta, \varphi)$ in Eq.~(\ref{scdisp}),
\begin{equation*} \mathcal{D}_N(\zeta) = \exp \left( i \frac{\tan^{-1} |\zeta| }{|\zeta|} \left( \Im(\zeta) J_x + \Re(\zeta) J_y\right)\right) .\end{equation*}
We have that \begin{equation*}\lim_{N \to \infty} \frac{\tan^{-1} | \zeta / \sqrt{2N} | }{| \zeta / \sqrt{2N} | }=1,\end{equation*} and hence
\begin{equation*} \begin{split} \lim_{N \to \infty}\mathcal{D}_N\left( \frac{\zeta}{\sqrt{2N}} \right) &  = e^{i\left(\Im(\zeta) \hat{x} - \Re (\zeta) \hat{p}  \right)}, \\ &  = \mathcal{D}( \Re(\zeta),\Im (\zeta)) , \end{split}\end{equation*}
which is the displacement operator for a field-mode given in Eq.~(\ref{entangledisp}). Furthermore via this contraction process we have that 
\begin{equation*} \lim_{N \to \infty} \left| \frac{ \zeta} {\sqrt{2N}} \right\rangle_N = \ket{\Re(\zeta),\Im(\zeta)}, \end{equation*}
where the right hand side is a field-mode coherent state, as defined in Eq.~(\ref{cohst})  \cite{radcliffe1971some,arecchi1972atomic,gazeau2009coherent, dooley2013collapse}.

\section*{References}

\bibliographystyle{apsrev}
\bibliography{MyLibrary} 

\begin{thebibliography}{60}
\bibitem[{not({\natexlab{b}})}]{noteA}
\bibinfo{note}{It is actually only necessary to be able to implement some basis
  changing gate directly on the qubits (such as the Hadamard gate) as diagonal
  single qubit gates can be implemented via the bus by making the second and
  fourth displacement in the sequence of Eq.~(\ref{rect}) unconditional
  displacements.}
\bibitem[{not({\natexlab{c}})}]{noteB}
\bibinfo{note}{In this definition, $2^{-1}$ is the multiplicative inverse of 2
  in $\mathbb{Z}(d)$ . $2^{-1}$ exists in $\mathbb{Z}(d)$ only if 2 and $d$ are
  coprime, that is only if $d$ is odd. However, as in the field-mode case, the
  phase factor is optional, and is included for notational simplicity and to
  emphasise the analogy with the field-mode case. Furthermore, the only
  displacements directly discussed will be of the form $\mathcal{D}(x,0)$ or
  $\mathcal{D}(0,p)$ so this technical detail will not be relevant and will not
  restrict the discussion to cases where $2^{-1}$ exists.}
\bibitem[{Not()}]{NoteD}
\bibinfo{note}{We may instead define the ancilla-register interaction to be of
  the form $C(\mathcal{D}_d(x,p),\mathcal{D}_d(-x,-p))$ to be even more
  obviously equivalent to the interaction in the qubus model. In the qubus
  model choosing between an interaction of the form
  $C(\mathcal{D}(x,p),\mathcal{D}(-x,-p))$ and one of the form
  $C\mathcal{D}(x,p)$ is essentially irrelevant (the choice taken herein is to
  be consistent with the literature) as they are locally equivalent and $x$ and
  $p$ can take any values. However in the qudit case the choice made here is
  more convenient as it allows lower dimensions of qudit for a set required
  smallest phase gate.}
\bibitem[{not()}]{noteC}
\bibinfo{note}{The phase space $\mathbb{Z}(d) \times \mathbb{Z}(d)$ only has a
  rigorous geometry when it can be labelled by a finite field (which is only
  possible when $d$ is an integer power of a prime). These technicalities are
  not relevant here as we may consider the phases as being geometric in origin
  in analogy to the phases created by displacements of a field mode.}
\end{thebibliography}

\end{document}